Timing of thermal metamorphism in CM chondrites: implication for Ryugu and Bennu future sample return

Elsa Amsellem[1], Frédéric Moynier[1,2], Brandon Mahan[1,3], Pierre Beck[2,4]

[1] Université de Paris, Institut de physique du globe de Paris, CNRS, F-75005 Paris, France
[2] Institut Universitaire de France, Paris, France.
[3] Department of Earth and Planetary Sciences, Macquarie University, Sydney, New South Wales 2109, Australia
[4] Institut de Planétologie et d'Astrophysique de Grenoble, Univ. Grenoble Alpes, CNRS, CNES, 38000 Grenoble, France

Abstract

Carbonaceous chondrites are often considered potential contributors of water and other volatiles to terrestrial planets as most of them contain significant amounts of hydrous mineral phases. As such, carbonaceous chondrites are candidate building blocks for Earth, and elucidating their thermal histories is of direct important for understanding the volatile element history of Earth and the terrestrial planets. A significant fraction of CM type carbonaceous chondrites are thermally metamorphosed or "heated" and have lost part of their water content. The origin and the timing of such heating events are still debated, as they could have occurred either in the first Myrs of the Solar System via short-lived radioactive heating, or later by impact induced heating and/or solar radiation. Since Rb is more volatile than Sr, and some heated CM chondrites are highly depleted in Rb, a dating system based on the radioactive decay of $^{87}$Rb to $^{87}$Sr ($\lambda^{87}$Rb = 1.393 x 10$^{-11}$ yr$^{-1}$) could be used to date the heating event relating to the fractionation of Rb and Sr. Here, we have leveraged the $^{87}$Rb/$^{87}$Sr system to date the heating of five CM chondrites (PCA 02012, PCA 02010, PCA 91008, QUE93005 and MIL 07675). We find that the heating events of all five meteorites occurred at least 3 Ga after the formation of the Solar System. Such timing excludes short-lived radioactive heating as the origin of thermal metamorphism in these meteorites, and relates such heating events to ages of collisional families of C-type asteroids.

# Introduction

Heated carbonaceous chondrites are meteorites that were thermally metamorphosed during one or several heating events subsequent to aqueous alteration for most of them (e.g., Nakamura, 2005, Alexander et al., 2012, 2013; Beck et al., 2014). Post aqueous alteration thermal metamorphism in CM chondrites was observed for the first time in samples collected in Antarctica, Yamato 793321 and Belgica 7904, because of the presence of an "intermediate phase" with "transformed" phyllosilicates, indicating matrix heating at a later stage than aqueous alteration (Akai, 1988). The matrix of typical CM chondrites is composed largely of the hydrous minerals serpentine (Mg-Fe silicate) and tochilinite (Fe sulphide), however most of the heated CM meteorites contain Fe-rich olivine, low Ca pyroxene and troilite (Fe sulphide) consistent with dehydration of the matrix minerals (Nakamura, 2005).

Heated meteorites show depletion in water content (Garenne et al., 2014), C (relative to H) (Alexander et al., 2013), trapped noble gases (Nakamura et al., 2006) and in highly and moderately volatile or labile elements (Rb, Cs, Se, Ag, Te, Zn, In, Bi, Tl; e.g., Wang and Lipschutz, 1998). The heating origin for these features is confirmed by the absence or lower amount of phyllosilicates and their dehydrated matrix compared to typical CM chondrites (e.g. Nakato et al., 2013), as well as higher organic maturity (Quirico et al., 2013, 2018).

Evaporation has been demonstrated as the origin for volatile element loss in these meteorites through stable isotopic measurements of Zn and Rb (Mahan et al. 2018, Pringle and Moynier 2017). Mahan et al. (2018) analysed different heated CMs for Zn isotopic composition and found that they are enriched in the heavier isotopes of Zn, most notably for PCA 02010 and PCA 02012, suggesting evaporation in an open system as the main culprit for Zn (and other volatile element) loss. Similarly, Pringle and Moynier (2017) reported high $^{7}Rb/^{85}Rb$ ratios associated with low Rb concentration for the same CMs, implying volatile loss during evaporation consistent with Rb fractionation under partial isotope equilibration between the evaporated gas and the residue. Cumulatively, these results lead to only three possible heating sources, either during the first Myrs of the solar system via short-lived radioactive heating, or later by impact induced heating or solar radiation when approaching the Sun. Presently, the main argument in favour of short duration heating, i.e. not from radioactive decay, is based on the absence of Fe-Mg diffusion into chondrules of the heated CM PCA 02012, as longer heating would lead to diffusion patterns (Nakato et al., 2013), however this argument does not remark on the actual dates of events. Determining the timing of the heating events can bring resolution to this longstanding issue.

Here, we report Rb-Sr model ages of 5 heated CM chondrites (PCA 02012, PCA 02010, PCA 91008, QUE93005 and MIL 07675). The $^{87}Rb$-$^{87}Sr$ radioactive decay system ($\lambda^{87}Rb = 1.393 \times 10^{-11}$ yr$^{-1}$; Nebel et al. 2011) is well suited to date volatile depletion events (e.g Papanastassiou and Wasserburg, 1969; Moynier et al., 2012; Hans et al., 2013). This is because Rb is relatively volatile (50% condensation temperature, $Tc_{Rb}$, of 800 K, Lodders 2003) while Sr is relatively refractory ($Tc_{Sr}$= 1404 K) implying a chemical fractionation of Rb and Sr during a volatilisation process. Heated CM chondrites have lost up to 90% of their Rb during thermal metamorphism leading to extremely low Rb/Sr ratios. If we assume a single stage fractionation model in which the Rb/Sr is fractionated from an unheated CM (for example equivalent to the CM chondrite Murchison), a model age of the volatilization event can be estimated from the $^{87}Sr/^{86}Sr$ ratio and the Rb/Sr ratio of the heated CM (Figure 1). Figure 1 presents the different evolution trends of the $^{87}Sr/^{86}Sr$ for a non-heated chondrite (trend without perturbation) and heated chondrites (the evolution is changed after the heating event where a part of Rb is lost from the system). We find that the ages of Rb/Sr fractionation are relatively recent (<2 Ga), implying that the CM were heated after the formation of their parent bodies and therefore not due to $^{26}Al$ radioactive decay. A possible heat source at this later stage could be represented by impacts. We therefore compare the epoch of meteorite heating events with the ages of some known asteroid collisional families, involving asteroids belonging to the spectroscopic C-types, which have been associated to carbonaceous chondrites.

2) Material and method

This section presents briefly the meteorite samples including their heating characteristics and presents the method used to estimate the timing of the heating events. This method leans on the Rb/Sr radioactive decay. During a heating event, the more volatile elements (e.g., Rb) are lost from the system unlike the more refractory elements (e.g., Sr), this leads to a fractionation between Rb and Sr. The loss of Rb in the asteroid parent body would change the evolution of the production of the radiogenic $^{87}Sr$ (represented by $^{87}Sr/^{86}Sr$). Comparing the heated chondrites with non-heated chondrites enables us to estimate the timing of this fractionation, corresponding to the heating event.

2.1) Samples and descriptions

A set of USGC standards were analysed as method validation: BIR-1a (Icelandic basalt); STM-2 (Table Mountain Syenite); AGV-2 (Guano Valley Andesite) and BCR-2 (Columbia River Basalt). For the CM chondrites, five Antarctic samples (PCA 02012, PCA 02010, PCA 91008, QUE93005 and MIL 07675) were chosen to represent a variety of thermally metamorphosed CMs. They are all considered as heated chondrites but the degree of heating varies between them.

PCA 02012 was first recognised as a thermally metamorphosed CM in the study of Nakato et al. (2013). The matrix is dehydrated and presents a high organic maturity but the elemental composition is close to typical CM2 chondrites. They estimate the temperature of the heating event to be approximately 900°C, as this meteorite compared with experimentally heated CM chondrites from the study of Nakato et al. (2008) share similar features. PCA 02012 represents a stage IV (strongly heated) according to the heating stage classification of Nakamura et al. (2005).

Stable isotopic compositions of PCA 02012 and PCA 02010 have been perfomed for Rb, Zn, H, (C and N) isotopes (Alexander et al., 2013; Mahan et al., 2018; Pringle and Moynier, 2017), with Rb and Zn in particular suggesting heating and volatile loss in an open system. These two chondrites have lost between 40% and 90% of Rb during one or several events (Pringle and Moynier, 2017) confirming the temperature of the heating source estimated, since the half condensation temperature of Rb is Tc= 800K (Lodders et al., 2003).

PCA 91008 shows depletion in most thermally mobile trace elements suggesting loss during heating (Wang and Lipschutz, 1998) and was classified as heating stage III containing poorly crystalline dehydrated phyllosilicate phase in the matrix (Nakamura et al., 2005). PCA 91008 shows evidence of terrestrial weathering by the analysis of the matrix and chondrules mineralogy and their textures (Tonui et al., 2014). Conventional stable isotopic compositions ($\delta D$, $\delta^{18}O$, $\delta^{13}C$ and $\delta^{15}N$) were obtained by Alexander et al. (2010, 2013). Transmission infrared spectra were obtained on PCA 02010 and PCA 91008 to reveal dehydration trends with phyllosilicates replaced by olivine (Beck et al., 2014), confirming earlier results of Tonui et al. (2002) and Quirico et al. (2011).

Loss of hydrogen is estimated for the three PCAs suggesting a dehydration event (Garenne et al., 2014). Furthermore, the content of C and H of the three meteorites relative to the isotopic composition of H reveals a severe heating process to explain the loss of C preferentially relative to H.

QUE 93005 was studied for petrological observations, aqueous alteration processes and O isotopic composition and was not viewed as thermally metamorphosed in these studies especially because it contain hydrated minerals (Rubin et al., 2012; Rubin et al., 2007; Clayton and Mayeda, 1999), however, it was classified as a heated CM (Alexander et al. 2012, 2013) because of its low hydrogen abundance correlating with H isotopic composition similar with other heated CM including MIL 07675 and PCA 91008. In addition, Raman spectra of QUE 93005 are similar to the three PCA studied here (Quirico et al., 2013). The analysis of insoluble organic matter's composition and structure shows evidence of thermal event but the presence of hydrated minerals argues against a thermal process. Quirico et al. (2018) have suggested that QUE 93005 experienced retrograde aqueous alteration, following the thermal event, which erased dehydration signatures.

The five chondrites contain carbonates that were analysed by Alexander et al. (2015) for C and O isotopic composition. The three PCA and MIL 07675 show the lowest carbonate content and the lowest $\delta^{13}C$ in carbonate compared to others CMs revealing a heating event.

2.2) Sample preparation

Rb-Sr chronology method requires the precise calculation of concentrations of Rb and Sr in the bulk samples to obtain the $^{87}$Rb/$^{86}$Sr ratio and the determination of the $^{87}$Sr/$^{86}$Sr ratio. The concentration of Rb and Sr is determined by isotope dilution technique that is based on adding a tracer enriched in a single isotope ($^{84}$Sr and $^{87}$Rb) to an aliquot of the samples and then determining the concentration via isotopic measurement. The optimum amount of spike added is determined by minimizing the error propagation on the calculated sample concentration. Previously calibrated spikes were added as solutions to the powdered sample prior to digestion in order to let spike and sample equilibrate, and spiked and un-spiked samples were digested in two steps. The powders were dissolved in a 3:1 mixture of concentrated HF and concentrated $HNO_3$ and placed on a hot plate for 2 days at 120°C. The samples were evaporated to dryness and concentrated hydrochloric acid was added during 1 day at 120 °C. The $^{87}$Sr/$^{86}$Sr ratio was determined on the unspiked samples following purification by ion exchange chromatography using 200 µL of pre-cleaned Sr-specific resin (Eichrom, 20-50 µm), following a protocol previously described (Amsellem et al., 2017). Briefly, the spiked samples were purified in order to elute both Rb and Sr. Rubidium was first purified on a pre-cleaned AG50X12 cation exchanged resin following one step of purification scheme following Pringle and Moynier (2017). The 1.8mL resin was conditioned with 3N HCl and the sample was loaded in 3N HCl. The matrix was eluted in 4mL of 3N HCl and the Rb was eluted with 6mL of 3N HCl. The matrix was collected to separate Sr from the other elements using the Sr-specific resin afterwards. The resin was conditioned with 3N $HNO_3$ and the sample was loaded in 3N $HNO_3$, Sr was eluted with Milli-Q $H_2O$ water after the elution of the matrix with 3N $HNO_3$. Two different powders of PCA 02012 were made through the entire chemistry process from the digestion to the measurement.

2.3) Measurements

Measurements of the Sr (both for natural isotopic composition and isotope dilution) and Rb (for isotope dilution) isotopic composition of the samples were performed in medium resolution mode on a Thermo Fisher Neptune Plus MC-ICP-MS located at the Institut de Physique du Globe de Paris (IPGP), during three different sessions. Samples were introduced using a cyclonic spray chamber. The intensities of $^{82}$Kr, $^{83}$Kr, $^{84}$Sr, $^{85}$Rb, $^{86}$Sr, $^{87}$Sr and $^{88}$Sr were collected using Faraday cups. We performed on-peak-zeros prior to each sequence and frequently during the sequence to monitor Kr interferences on mass $^{84}$Sr and $^{86}$Sr.

For the Sr natural isotopic measurement, the mass-dependent fractionations (natural and instrumental), on the $^{87}Sr/^{86}Sr$ ratio were corrected through internal normalization to a $^{86}Sr/^{88}Sr$ value of 0.1194. The accuracy of the $^{87}Sr/^{86}Sr$ ratio were confirmed by analysing the NIST SRM987 standard giving a value of 0.71029 ± 0.00004 which falls in between the certificate value (0.71034) and the commonly accepted value (0.71026) (Balcaen et al., 2005). For the isotope dilution measurements of Sr and Rb, the $^{84}Sr/^{86}Sr$ ratio and $^{87}Rb/^{85}Rb$ ratio were measured respectively.

3) Results

The $^{87}Sr/^{86}Sr$ and $^{87}Rb/^{86}Sr$ ratios are presented in Table 1 for the standard rocks and the five bulk chondrites. The Rb and Sr concentration of the standards are compared with the literature data (USGS Certificate of Analysis and published values) in Figure 2a and 2b. The Sr concentration are similar with the literature within error except for STM-2 where there is 10% of difference with USGS data but is within error compared to Schudel et al. (2015). The Rb concentrations are similar within error compare to the literature. The $^{87}Sr/^{86}Sr$ of the standards are compared with the literature in Figure 2c. The Rb concentration of BIR-1a is not presented, as an incorrect amount of spiked solution was added in the sample leading to an incorrect value.

Rubidium concentrations of the five heated chondrites are all lower than typical CM (Murchison; ~1.6 ppm; Mittlefehldt and Wetherill, 1979 and Pringle and Moynier 2017, Figure 3), and for PCA 02010 (Rb= 0.24ppm) it is consistent with concentrations obtained previously during isotopic measurements after intensive Rb chemical purification and by comparing the intensity of the Rb of the sample with a standard of known concentration (Rb= 0.2ppm, Pringle and Moynier, 2017). PCA 02010 and PCA 02012 chondrites show similar Sr concentration with typical CM (Murchison; ~10 ppm; Mittlefehldt and Wetherill, 1979) whereas PCA 91008, QUE 93005 and MIL07675 are depleted in Sr compared to typical CM. QUE 93005 is depleted in Rb, which could be related to evaporation of Rb without affecting the hydrated minerals. The $^{87}Sr/^{86}Sr$ ratio of the heated chondrites (from 0.72438 ± 0.00009 to 0.73010 ± 0.00014) are within the range but slightly lower of typical CM chondrites (from 0.72765 ± 0.00005 to 0.73240 ± 0.00095; Mittlefehldt and Wetherill, 1979). The $^{87}Rb/^{86}Sr$ range from 0.043 ± 0.001 to 0.326 ± 0.003 and are lower compared to the non-heated CM chondrites (~0.4; Mittlefehldt and Wetherill, 1979). The two different powders of PCA 02012 show slight differences in Rb concentration and in $^{87}Sr/^{86}Sr$ ratio reflecting the heterogeneity of the sample.

The chondrites from this study are Antarctic finds and are thus subjected to alteration. However, the effect of alteration on Sr would not affect the results as we internally normalize the isotope ratios to $^{86}Sr/^{88}Sr$, correcting for mass dependent fractionations such as created by alteration processes. Furthermore, there is no clear correlation between Rb and weathering grades (data from Alexander et al., 2012) and Rb abundances in Antarctic CM chondrites and non-Antarctic ones are similar, suggesting that Rb is not affected by alteration for these samples.

4) Discussion

The timing of the Rb/Sr fractionation can be estimated using a 2-stage model in which the first stage is the evolution of a CM chondrite reservoir formed at $T_0$ (4.567 Ga, Connelly et al., 2012) and the second stage represents the Rb/Sr fractionation event that occurred at time $T_H$ using the following equations, where T corresponds to the present time and $\lambda$ the decay constant rate:

$$\left(\frac{^{87}Sr}{^{86}Sr}\right)_H^T = \left(\frac{^{87}Sr}{^{86}Sr}\right)_H^{T_H} + \left(\frac{^{87}Rb}{^{86}Sr}\right)_H^T \times (e^{\lambda(T-T_H)} - 1) \qquad \text{Eq.1}$$

$$\left(\frac{^{87}Sr}{^{86}Sr}\right)_H^{T_H} = \left(\frac{^{87}Sr}{^{86}Sr}\right)_{NH}^{T_H} = \left(\frac{^{87}Sr}{^{86}Sr}\right)_{NH}^{T_0} + \left(\frac{^{87}Rb}{^{86}Sr}\right)_{NH}^{T_H} \times (e^{\lambda(T_H-T_0)} - 1) \qquad \text{Eq.2}$$

$$T_H = -\frac{1}{\lambda}\ln\left(\frac{\left(\frac{^{87}Sr}{^{86}Sr}\right)_H^T - \left(\frac{^{87}Sr}{^{86}Sr}\right)_{NH}^T}{\left(\frac{^{87}Rb}{^{86}Sr}\right)_H^T - \left(\frac{^{87}Rb}{^{86}Sr}\right)_{NH}^T}\right) \qquad \text{Eq.3}$$

Equation 1 represents the radioactive decay equation showing the production of the radiogenic $^{87}Sr$ over the stable $^{86}Sr$ of a heated chondrite from the present time T to $T_H$ (timing of the heating event). Equation 2 represents the evolution of the $^{87}Sr/^{86}Sr$ for a non-heated chondrite and heated chondrite (which is the same) from $T_o$ to $T_H$. Here, the assumption is that the initial $^{87}Sr/^{86}Sr$ and $^{87}Rb/^{86}Sr$ ratios of the heated chondrites (subscript H) were similar to the ones of the non-heated chondrites (subscript NH). The timing of the heating event (Equation 3) is deduced by Equations 1 and 2.

The $^{87}Sr/^{86}Sr$ and $^{87}Rb/^{86}Sr$ ratios of non-heated chondrites is sligthly variable (Mittlefehldt and Wetherill, 1979; Kaushal and Wetherwill, 1970) and therefore introduce an uncertainty in the estimation of the heating age as the calculated timing will depend on which non-heated CM is chosen as reference. Here, we decided to use the two non-heated CMs that represent the most primitive chondrites in term of heating event (Quirico et al., 2018) and with the most extreme $^{87}Rb/^{86}Sr$ and $^{87}Sr/^{86}Sr$ ratio (Mighei and Murchison, Mittlefehldt and Wetherill, 1979; Kaushal and Wetherwill, 1970) in order to obtain the most conservative errors. This calculation has been done for each of the 5 heated CM chondrites and the results are presented in Table 2 and Figure 4. The ages range between 1.54 Ga ± 0.48 Ga and 607 ± 614 Ma ago, indicating that all heating events herein were younger than 2.02Ga.

This represents the first dating of the thermal metamorphism of CM chondrites and indicates that these events occurred relatively late in Solar System history. This observation excludes radioactive decay of $^{26}Al$ as the source of heating for CM metamorphism as it would have all decayed away a long time before (~less than 10Myrs after Solar System formation, e.g., Lee et al., 1977). This result alone provides a major step in our understanding of the mechanisms of heating for metamorphism.

The two alternative sources of heating that could occur later in the Solar System history are solar heating and heat released during impacts. Solar radiation on asteroids could occur when a body approaches the Sun. The two recent CM chondrite falls Sutters's Mill and Maribo both had perihelion distances below 0.5 A.U. (Jenniskens et al., 2012). This implies an equilibrium temperature around 550 K at the subsolar point (at perihelion, using an albedo of 0.01), which is a minima since the orbital evolution prior to delivery on Earth may have encompassed even closer approaches to the Sun (Marchi et al., 2009). If the heating is due to solar radiation, one will expect young ages for the heating event, of the order of the Cosmic-Ray Exposure (CRE) obtained on carbonaceous chondrites (typically < 2 Myrs, Eugster et al., 2006). The age determined for PCA 91008 is much older that these CRE ages as well as typical transfer that can be estimated from solar system dynamics (Morbidelli and Gladman, 1998). In the case of the 4 other samples, the error on the ages does not allow us to strictly rule out solar radiation.

Our data cannot settle this issue but looking at petrographic studies of PCA 02012 suggest that the heating event may have lasting less than 10 seconds and at ~900°C (Nakato et al., 2013). This is based on the Fe-Mg interdiffusion between chondrules, the thermal maturation of the organic matter, and the texture of the matrix. This short duration is in agreement with the expected duration of impact metamorphism (Beck et al., 2005).

In addition, QUE 93005 seems not to have been homogeneously heated. It has kept some hydrated minerals while lost >50% of its Rb. The heterogeneous temperature distribution may be representative of impact heating, as depending on the porosity of the sample, a higher post-shock temperature is induced and may be heterogeneously distributed (Schmitt et al.,

1994). This sample is the least heated chondrite from this study and has also a very weak evidence of shock (shock stage S1, Rubin et al., 2012). All together, the late ages, the petrographic features and the heterogeneous heating suggests that impact heating is the most likely source of heating to explain the thermal metamorphism of CM chondrites. C-type asteroids are the most common asteroids in the main belt and are the closest spectral analogues to carbonaceous chondrites (e.g., Johnson and Fanale, 1973; Chapman et al., 1975). We will now discuss the relation between the impacts at the origin of the thermal metamorphism observed in the chondrites and the collisional history of C-type asteroids.

Shock experiments conducted on Murchison have showed that in order to dehydrate phyllosilicates and form Fe-sulfide, olivine and pyroxene, an impact pressure of at least 30 GPa is needed (Tomeoka et al., 2003). This corresponds to a velocity impact of ~2 km.s$^{-1}$ for a metallic impactor (Tyburczy et al., 1986), for a silicate one the velocity is expected to be higher which would be close to the mean value of 5 km.s$^{-1}$ in the asteroid belt (Farinella and Davis, 1992). Therefore, physical conditions required to heat the CM chondrites are on par with those encountered in the asteroid main-belt (e.g., Wasson et al., 1987; Rubin, 1995). Furthermore, the similarity in the reflectance spectra between naturally heated CMs, experimentally heated CMs and some type-C, G, B and F asteroids have already suggestively been linked with the same process: impacts in the asteroid belt (Hiroi et al., 1993). This is consistent with the recent results of reflectance spectra on Ryugu's surface from Hayabusa2 mission that are similar to thermally or shock-metamorphosed carbonaceous chondrites (Kitazato et al., 2019; Sugita et al., 2019). However, metamorphosed CI and CM chondrites have been suggested not to represent the surface material of asteroids B, C, Cb and Cg (BCG) types as the density expected for heated asteroids would be higher than those of Ch and Cgh types (hydrated type asteroids) which is not the case for BCG types asteroids (Vernazza et al., 2015). Thus, the association of heated chondrites with their asteroid counterparts is not clear.

There are three types of collision that might have played a role in setting the age of heated CM chondrites. The Rb loss might have occurred during different heating events. CMs are thought to have been compacted by a series of impact events (Lindgren et al., 2015), which would lead to different heating events. Our age would then represent a minimum age for these events.

i) First, we can think of "old collisions", related to the initial accretion phases of the Solar System, and the late heavy bombardment. In that case, we would expect the heating ages to be similar to the K/Ar ages of lunar samples, typically 3.8 Gyrs and older. This is not the case for the heated CM chondrites measured here. This excludes direct ejection of the meteorites from a large primordial C-type with an old and heavily cratered surface.

ii) Second, collisions in the asteroid belt are still occurring and the asteroid family represents the main source of meteorites (Vernazza, 2008; 2014) hence the young ages of the heating event of the CM chondrites could be due to collisions forming asteroid families. Collision ages of C-type asteroid families can be compared to the ages obtained here for CM chondrites to discern whether the collisional heating could be related to disruptive, family forming, impacts (Figure 4). The ages of the collisions that created different asteroids families are estimated by modelling the spreads of the orbital parameter of family members due to Yarkovsky effect (Brož et al., 2013; Bottke et al., 2015; Spoto et al., 2015; Milani et al., 2014, 2017; Paolicchi et al., 2018). Asteroid family Misa, Erigone, Naema, Astrid, Brasilia and 1993 FY12 (C-types, Carruba et al., 2013) have ages (259 ± 95 Ma, 224 ± 36 Ma, 206 ± 45 Ma, 156 ± 23 Ma, 150 ± 23 Ma, 143 ± 56 Ma and 83 ± 21 Ma; Paolicchi et al., 2018) similar within error to the heating ages estimated here for PCA 02012, PCA 02010, MIL 07675 and QUE 93005. Prokne, Hygiea, Euphrosyne, Klumpkea, Dora, Hoffmeister, Astrae (C-types, Carruba et al., 2013), New Polana and Eulalia (Walsh et al., 2013) asteroid family with ages of (1448 ± 348 Ma, 1347 ± 220 Ma, 1225 ± 304 Ma, 663 ± 154 Ma, 506 ± 116 Ma, 463 ± 110 Ma, 332 ± 67 Ma, 329 ± 50 Ma, 328 ± 71 Ma, 1.4 ± 0.15 Ga and 0.8 ± 0.1 Ga respectively; Paolicchi et al., 2018 and Bottke et al., 2015) are similar to PCA 91008, PCA 02012 and QUE 93005 (Figure 4). The two asteroids Prokne and Euphyrosyne have surface spectra similar to CM2 chondrites (Cloutis et al., 2011) and are thus good family candidates for PCA 91008, PCA 02012 and/or QUE 93005. New Polana and Eulalia asteroid family are thought to be at the origin of Bennu asteroid (Bottke et al., 2015). Bennu has similar thermal infrared spectral features as CM chondrites (Hamilton et al., 2019) and is visited by OSIRIS-REX project. Both families are ideally placed in the inner portion of the main belt to escape from the asteroid belt to Earth. While the uncertainties on the ages are large, this illustrates a certain consistency between astronomical observation and meteoritic record and suggest that the heating event at the origin of the thermal metamorphism of the heated CM may originate in the disruptive events at the origin of the asteroid families.

iii) Last, many near-Earth asteroids that have been visited by spacecrafts are rubble-piles, and that are explained today by catastrophic collisions (Michel et al., 2005). The rubble-pile forming collision could also have played a role in setting the Rb/Sr age of the heated CM chondrites. It is possible that family-forming collisions generate rubble-piles and that (ii) = (iii). From a modelling perspective it appears possible (Michel et al., 2001).

CRE ages have already been used to link weakly heated chondrites with their asteroid counterparts. Meier et al. (2016) compared CRE ages of Jbilet Winselwan (King et al., 2019) with the timing of the disruption of the parent body of Veritas family. The combination of several techniques may improve our understanding on the relation between meteorites and asteroids families in the asteroid belt.

# Conclusion

We developed a new model to date the heating event of aqueously altered and thermally metamorphosed chondrites using Rb-Sr chronology and applied it to five heated CM chondrites. The five chondrites were heated during the two last billion years excluding $^{26}$Al radioactive decay as the origin of the heating. These young ages more likely reflect collision forming events in the asteroid belt. Our model age results are consistent with the estimated ages of impacts of C-type asteroids from physical observations, suggesting that in the future we may be able to link a heated chondrite with its asteroid family by comparing ages.

Acknowledgements

FM acknowledges funding from the European Research Council under the H2020 framework program/ERC grant agreement #637503 (Pristine) and financial support of the UnivEarthS Labex program at Sorbonne Paris Cité (ANR-10-LABX- 0023 and ANR-11-IDEX-0005-02), and the ANR through a chaire d'excellence Sorbonne Paris Cité.

Figures

Figure 1

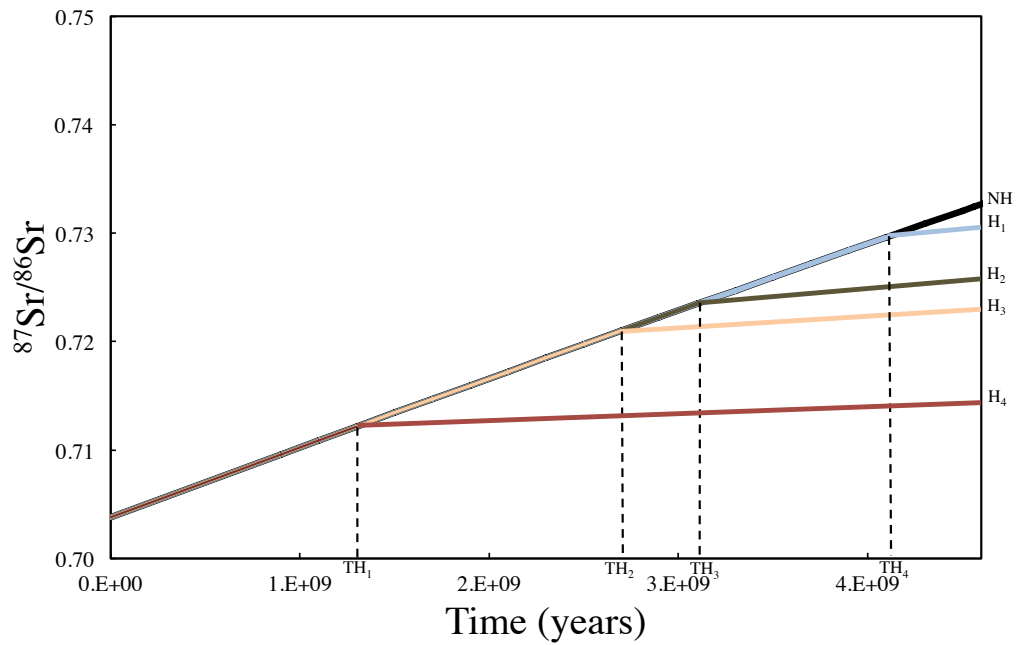

Figure 2a

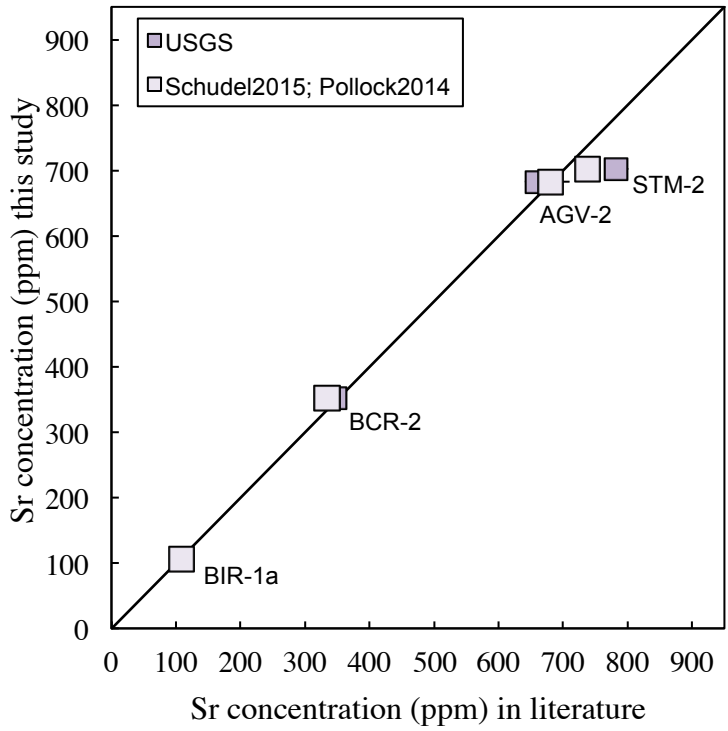

Figure 2b

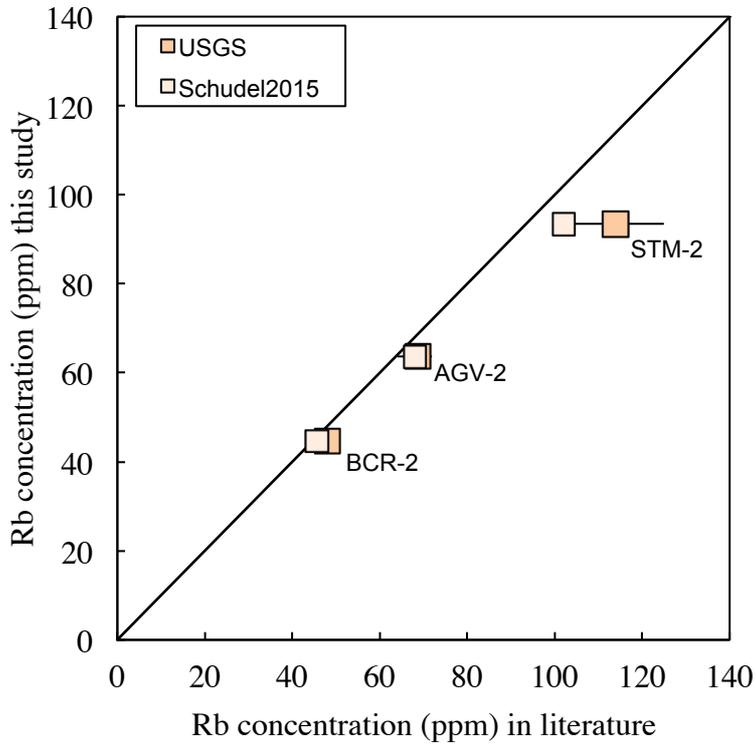

Figure 2c

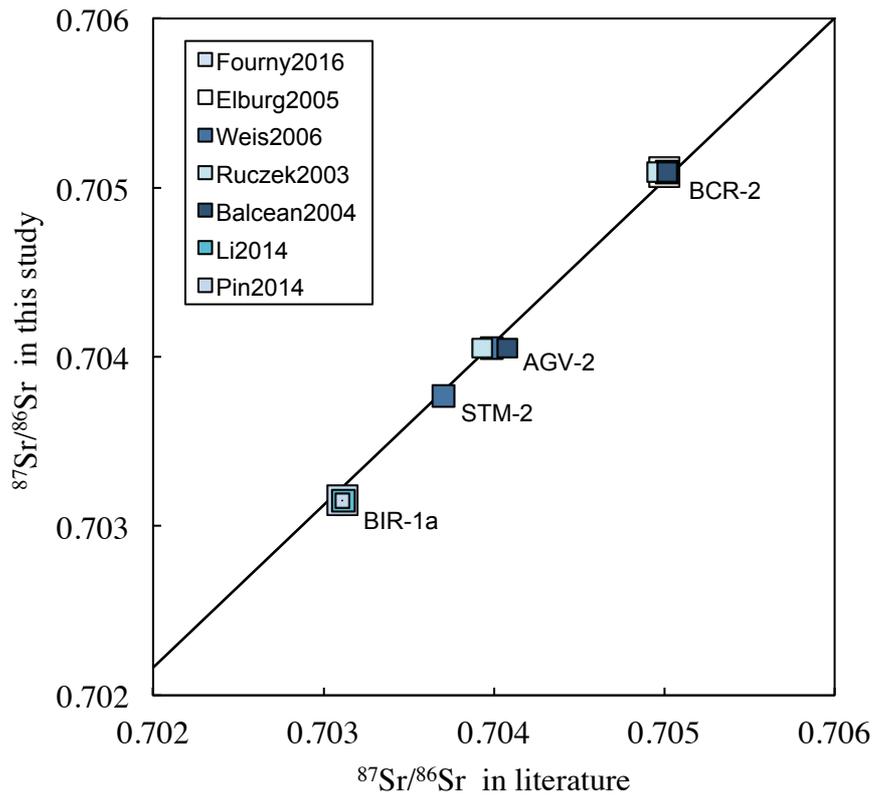

Figure 3

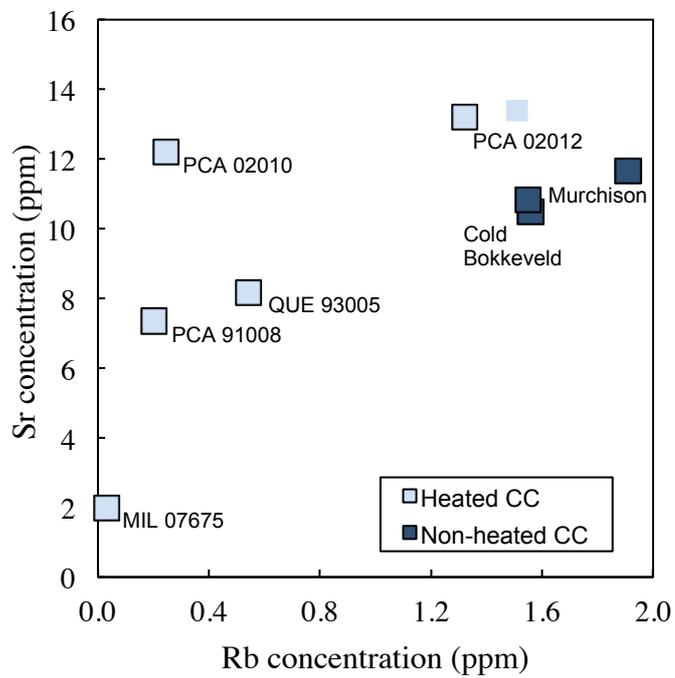

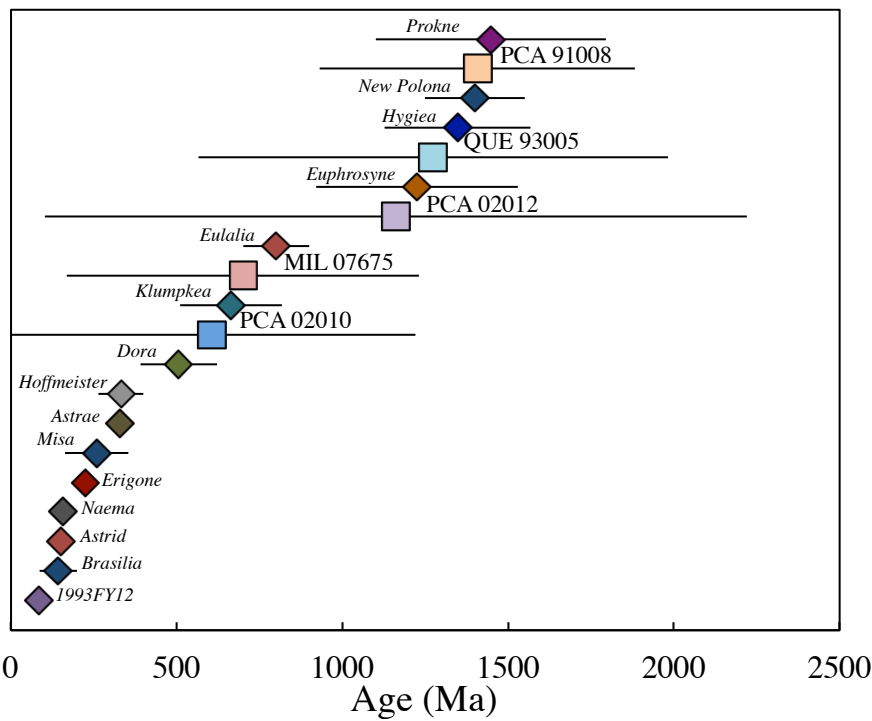

Figure 4

Captions

Figure 1: Fictive Evolution of the $^{87}Sr/^{86}Sr$ ratio of bulk chondrites ($H_x$) that occurred heating event at different time ($TH_x$) showing the use of Rb/Sr radiochronology to estimate the date of the heating event.

Figure 2: Sr (a) and Rb (b) concentrations and $^{87}Sr/^{86}Sr$ (c) ratios of terrestrials standards conducted in this study compare to the literature. The black line represents 1:1 slope.

Figure 3: Sr concentration of heated and non-heated CM chondrites relative to their Rb content. The light blue square symbol without an outline is the replicate of PCA 02012.

Figure 4: Estimation of the ages of heated CM (square symbols) compared to ages of collisional C-type asteroids family (diamond symbols) from Paolicchi et al. (2018).

**Table 1**
Rb and Sr concentration, $^{87}Sr/^{86}Sr$ and $^{87}Rb/^{86}Sr$ ratios of terrestrial basalts and heated CM chondrites

| Sample | $^{87}Sr/^{86}Sr$ | 2se[1] | [Sr] *ppm* | error[2] | [Rb] *ppm* | error[2] | $^{87}Rb/^{86}Sr$ |
|---|---|---|---|---|---|---|---|
| ***Heated CM chondrites*** | | | | | | | |
| PCA 02010 | 0.73010 | 0.00014 | 12.21 | 0.08 | 0.245 | 0.0027 | 0.057 |

| Sample | ⁸⁷Sr/⁸⁶Sr | 2se | Sr (ppm) | 2se | Rb (ppm) | 2se | ⁸⁷Rb/⁸⁶Sr |
|---|---|---|---|---|---|---|---|
| PCA02012 #1 | 0.72984 | 0.00009 | 13.21 | 0.10 | 1.320 | 0.0062 | 0.289 |
| PCA02012 #1 | 0.72968 | 0.00003 | 13.39 | 0.09 | 1.509 | 0.0152 | 0.326 |
| *Average* | 0.72976 | 0.00022 | 13.30 | 0.26 | 1.415 | 0.27 | 0.308 |
| PCA 91008 | 0.72438 | 0.00009 | 7.36 | 0.05 | 0.205 | 0.0013 | 0.079 |
| QUE 93005 | 0.72801 | 0.00002 | 8.17 | 0.06 | 0.542 | 0.0023 | 0.191 |
| MIL 07675 | 0.72898 | 0.00006 | 2.00 | 0.03 | 0.033 | 0.0004 | 0.043 |
| **Terrestrial basalts** | | | | | | | |
| BIR-1a | 0.70315 | 0.00004 | 105.50 | 0.24 | - | - | - |
| AGV-2 | 0.70405 | 0.00004 | 682.79 | 0.56 | 63.67 | 0.41 | 0.270 |
| STM-2 | 0.70377 | 0.00004 | 701.66 | 0.52 | 93.47 | 0.49 | 0.385 |
| BCR-2 | 0.70509 | 0.00006 | 352.20 | 0.45 | 44.58 | 0.05 | 0.366 |

[1] 2se= 2 x standard deviation /√n
[2] the error is estimated using a propagation law of error

**Table 2**
Age estimation of the heating event of five heated chondrites in Ma

| | Minimum value *Murchison* | Maximum value *Mighei* | Mean value | Standard error | Heating Stage |
|---|---|---|---|---|---|
| **PCA 02010** | -7 | 1221 | **607** | **614** | |
| **PCA02012 #1** | 102 | 2220 | **1161** | **1059** | III/IV |
| *PCA02012 #2* | 254 | 2580 | **1417** | **1163** | |
| **PCA 91008** | 1068 | 2019 | **1544** | **475** | III |
| **QUE 93005** | 566 | 1982 | **1274** | **708** | II |

| MIL 07675 | 168 | 1231 | **699** | **531** | - |